\documentclass[prb,twocolumn,showpacs,preprintnumbers,amsmath,amssymb]{revtex4}
\usepackage[normalem]{ulem}

\usepackage{graphicx}
\usepackage{color} 
\usepackage{amsmath}

\begin{document}

\title{Spin-glass phase transition and behavior of nonlinear susceptibility in 
the Sherrington-Kirkpatrick model with random fields}

\author{C. V. Morais$^1$, F. M. Zimmer$^2$,  M. J. Lazo$^3$, S. G. Magalh\~aes$^4$, 
F.~D. Nobre$^5$}

\affiliation{$^1$Instituto de F\'{i}sica e Matem\'atica, Universidade Federal 
de Pelotas, 96010-900  Pelotas, RS, Brazil}
\email{carlosavjr@gmail.com}

\affiliation{$^2$Departamento de Fisica, Universidade Federal de Santa Maria,
97105-900 Santa Maria, RS, Brazil}%

\affiliation{$^3$Programa de P\'os-Gradua\c{c}\~{a}o em F\'{\i}sica - Instituto de 
Matem\'atica, Estat\'{\i}stica e F\'{\i}sica, 
Universidade Federal do Rio Grande, 
96.201-900, Rio Grande, RS, Brazil }

\affiliation{$^4$Instituto de Fisica, Universidade Federal do Rio Grande do Sul, 91501-970
Porto Alegre, RS, Brazil}

\affiliation{$^5$ Centro Brasileiro de Pesquisas F\'{\i}sicas  and 
National Institute of Science and Technology for Complex Systems, 
Rua Xavier Sigaud 150, 
22290-180, Rio de Janeiro, RJ, Brazil}

\begin{abstract}

The  behavior of the nonlinear susceptibility $\chi_3$ and its relation to the  
spin-glass transition temperature $T_f$, in the presence of random fields, are investigated. 
To accomplish this task, the  Sherrington-Kirkpatrick model 
is studied through the replica formalism, within a one-step 
replica-symmetry-breaking procedure. 
In addition, the dependence of the Almeida-Thouless eigenvalue 
$\lambda_{\rm AT}$ (replicon) 
on the random fields is analysed. Particularly, in absence of random fields, the temperature 
$T_f$ can be traced  by a divergence in the spin-glass susceptibility  
$\chi_{\rm SG}$, which 
presents a term inversely proportional to the replicon $\lambda_{\rm AT}$.  
As a result of a relation between $\chi_{\rm SG}$ and  $\chi_3$, the latter also presents a 
divergence at  $T_f$, which comes as a direct consequence 
of $\lambda_{\rm AT}=0$ at $T_f$. However, our results show that, in the 
presence of random fields, $\chi_3$ presents a rounded maximum at a temperature 
$T^{*}$, which does not coincide with the spin-glass transition temperature $T_f$
(i.e., $T^* > T_f$ for a given applied random field). Thus, the maximum value of $\chi_3$ at $T^*$ 
reflects the effects of the random fields in the paramagnetic phase, 
instead of the non-trivial ergodicity breaking associated with the spin-glass phase 
transition. It is also shown that $\chi_3$ still maintains a dependence on 
the replicon $\lambda_{\rm AT}$, although in a more complicated way, 
as compared with the case without random fields.
These results are discussed in view of recent observations in 
the LiHo$_x$Y$_{1-x}$F$_4$ compound. 

\vskip \baselineskip


\end{abstract}
\maketitle


\section{Introduction}

The presence of disorder in spin systems 
represents a permanent source of 
challenging problems, due to the richness of physical properties that emerge 
from the interplay between disorder and many-spin interactions. 
Random fields (RFs) and spin glass (SG) models are important examples of such 
richness~\cite{Dotsenko,Nishimori,Young}. 
Furthermore, the combination of 
these two highly nontrivial manifestations of disorder leads to a fascinating area of 
research in spin systems, which is not only a theoretical possibility. 
Actually, they can be found in diluted Ising-like antiferromagnets, like  
Fe$_x$Zn$_{1-x}$F$_2$ and Fe$_x$Mg$_{1-x}$Cl$_2$~\cite{Belanger}.
Additionally, recent investigations have suggested the diluted 
Ising-like dipolar ferromagnetic compound LiHo$_x$Y$_{1-x}$F$_4$  
as a new candidate presenting these two types of disorder, 
bringing novel interesting and controversial 
issues~\cite{Schechter1,Laflorencie06,Gingras06,Gingras11,jonsson07,%
anconatorresprl2008,mydosh2015, Fernandez2010, Alonso2015}.
For instance, the LiHo$_{0.167}$Y$_{0.833}$F$_4$ compound, in the absence 
of an applied transverse field $B_t$, displays $\chi_3$ (the lowest term of the 
non-linear susceptibility $\chi_{nl}$) 
with a sharp peak at the freezing temperature $T_f$, which 
resembles a conventional second-order SG phase 
transition~\cite{wada,suzuki,Chalupa77}.
On the other hand,  the sharp peak of $\chi_3$ becomes increasingly 
rounded when 
$B_t$ is enhanced, being located at the temperature $T^*$, which 
is lower than $T_f$  obtained in absence of $B_t$~\cite{Wu91}.

The suggestion that an effective longitudinal RF can be induced by the interplay 
of a transverse applied field $B_t$ with the off-diagonal terms of the dipolar 
interactions in the LiHo$_x$Y$_{1-x}$F$_4$~\cite{Laflorencie06,Schechter1,Gingras06,Gingras11} 
brought a new push to clarify the controversies of the experimental behavior 
of $\chi_3$ and, therefore, the meaning of $T^*$, i.e., whether or not it is a 
true SG transition temperature. In the droplet picture used in 
Refs.~\onlinecite{Schechter1} and \onlinecite{Laflorencie06}, the presence of a RF $h_i(B_t)$, 
induced by the uniform transverse field, suppresses the SG transition, 
for the same reason that 
an uniform field does it in that picture~\cite{Fisher86}. On the other hand, 
within the mean-field Parisi's framework~\cite{Parisi80a,Parisi80b}, 
Tabei and collaborators~\cite{Gingras06}, using a quantum version of 
the Sherrington-Kirkpatrick (SK) model~\cite{SK} with additional off-diagonal 
interactions, longitudinal RF $h_i(B_t)$ and a  transverse field $\Gamma(B_t)$, 
succeeded in reproducing the $\chi_3$ experimental behavior. 
Indeed, this result is a strong evidence that the RF plays an important role in 
the LiHo$_x$Y$_{1-x}$F$_4$ compound. 
These authors also suggested that the SG quantum criticality is unlikely 
in this transverse field, induced longitudinal RF scenario;  
additionally, susceptibility measurements presented evidence of 
a canonical SG behavior~\cite{anconatorresprl2008,mydosh2015}. 
From that point of view, one can raise the question 
of what happens with the SG criticality in a regime where thermal fluctuations 
should be dominant as compared with the quantum ones? 
One possible consequence of the transverse field, induced longitudinal RF, 
is that the Almeida-Thouless (AT) 
line~\cite{Almeida78} can be suppressed, as suggested by numerical simulations 
in short-range-interaction SGs~\cite{Young04}.
However, previous studies using mean-field Parisi's framework  
have shown that the SK model with a RF does preserve the AT 
line~\cite{SNA,NNCC,ANC,MMN,Morais}. 
Consequently,  assuming that Parisi's mean-field theory is a valid framework to 
describe the SG problem with a transverse field, induced longitudinal RF, 
one can also raise the question of how the behavior of $\chi_3$ can be related 
with the AT line, when a RF is present in the SK model? 
One can expect that the answer to this question may also help 
to clarify the meaning of the temperature $T^*$.

Therefore, in this work we present a detailed investigation of the role of a RF in the  
behavior of $\chi_3$ and its relation to the AT line in the SK model within the 
mean-field Parisi's framework. In order to relate with experimental verifications on
the LiHo$_x$Y$_{1-x}$F$_4$ compound, we assume that $B_t$ is sufficiently small 
to assure that quantum fluctuations are negligible, but enough to guarantee that the 
effective field-induced RF $h_i(B_t)$ is still appreciable.   
When $h_i=0$ ($B_t=0$), it is known that $\chi_3$ is related with the 
SG susceptibility, 
\begin{equation}
\chi_{SG} = (\beta/N) \sum_{i,j} \left[ \left(\langle S_iS_j\rangle 
- \langle S_i \rangle \langle S_j \rangle  \right)^2 \right]_{\rm av}~, 
\label{eqchisg}
\end{equation}
where, as usual, $\langle .. \rangle$ and $ \left[ .. \right]_{\rm av}$ denote respectively, 
thermal averages and an average over the disorder. 
The SG susceptibility has a term inversely proportional to the AT eigenvalue 
$\lambda_{\rm AT}$, the so-called replicon~\cite{Binder}. 
Therefore, the diverging behavior, $\chi_3\propto (T-T_f)^{-\gamma}$, 
in the SG transition, is directly related with $\lambda_{\rm AT}=0$ 
at $T_f$, corresponding to the onset of replica-symmetry breaking (RSB).

However, the situation changes considerably when $h_i\neq 0$ 
(i.e., $B_t\neq 0$). For instance, the RF can induce directly the 
SG order parameter, but only in the replica-symmetric (RS)
approximation, since the RF 
is unable to produce any RSB.   
This result was demonstrated not only 
for infinite-ranged spin interactions~\cite{Pytte77},
but also for the Bethe lattice~\cite{Krzakala10}. 
As a consequence, the smooth behavior of  
the SG RS order parameter $q$ is not appropriate for
identifying a SG transition of the SK model in the presence of a random field;
however, such a transition may be related with the onset of RSB, 
associated with the replicon $\lambda_{\rm AT}=0$~\cite{SNA}. 
In spite of this, the derivative of $q$ with respect to the temperature increases
as one approaches $T_{f}$ from above; such an increase is responsible
for the rounded maximum in $\chi_3$ at a temperature $T^*$, 
which does not coincide with the SG transition temperature $T_f$
(i.e., $T^* > T_f$ for a given applied random field).  
Thus, the maximum value of $\chi_3$ at $T^*$ 
should reflect the effects of the RF 
inside the paramagnetic (PM) phase, 
instead of the non-trivial ergodicity breaking of the SG phase transition. 
Our  results also suggest that $\chi_3$ still maintains a dependence on the replicon 
$\lambda_{\rm AT}$, although in a much more complicated way, 
as compared with the case without the RF.

This paper is structured as follows. In the next section we define the model
and the analytical procedure to be used; then, we calculate $\lambda_{\rm AT}$,  
the order parameters within the one-step replica symmetry breaking 
(1S-RSB) scheme, the susceptibilities $\chi_1$, $\chi_3$, as well as 
the temperature $T^*$ in the presence 
of RFs, following both Gaussian and bimodal distributions.
In section III we discuss the numerical
solutions of the saddle-point equations for the order parameters and
susceptibilities. Finally, the last section is reserved to the conclusions.

\section{Model and susceptibilities}

Herein we consider the infinite-range-interaction spin-glass model,
defined by the following Hamiltonian
\begin{equation}
 H = - \sum_{(i,j)} J_{ij} S_i S_j - \sum_{i=1}^{N} h_{i} S_{i} - H_l \sum_{i=1}^{N} S_i, 
\label{eq1}
\end{equation}
where $S_{i}=\pm 1$, $H_l$ represents a uniform field, 
and the sum $\sum_{(i,j)}$ applies to all distinct pairs of spins.
The spin-spin couplings $\{ J_{ij} \} $ and the magnetic random fields $\{ h_i \}$ 
follow independent Gaussian probability distributions, 
\begin{equation}
P(X)=\left[\frac{1}{2\pi \sigma^{2}}\right]^{1/2} 
\exp\left[-\frac{X^2 - C}{2\sigma^{2}}\right],
\label{eq2a}
\end{equation}
where $X$ may represent either couplings or random fields; in the former case
one has $\sigma=J/\sqrt{N}$ and $C=J_{0}/N$, whereas in the later, 
$\sigma=\Delta$ and $c=0$. 
We also consider a bimodal probability distribution,  
\begin{equation}
\label{eq:phi}P(h_{i}) = p \ \delta (h_{i} - h_{0}) + (1-p) \ 
\delta (h_{i} + h_{0})~,  
\end{equation}
for the random fields $\{ h_i \}$. 
We follow closely the procedure used in Ref.~\onlinecite{SNA} to obtain the average free 
energy per spin, 
$f=-1/(\beta N) [ \mbox{ln} Z(\left\{J_{ij}\right\},\left\{h_i\right\}) ]_{J,h}$, 
where $Z(\left\{J_{ij}\right\},\left\{h_i\right\})$ represents the partition function 
for a given quenched distribution of random couplings and 
fields; moreover,  $[ . . ]_{J,h}$ denotes averages over these types of disorder,  
and $\beta=1/T$.
As usual, the replica method~\cite{Dotsenko,Nishimori,Binder} is applied; thus, 
\begin{equation}
-\beta f=\lim_{N\rightarrow\infty}\lim_{n\rightarrow 0}\frac{1}{Nn}
\left([ Z(\left\{J_{ij}\right\},\left\{h_i\right\})^{n} ]_{J,h}-1\right),\\
\label{eq4}
\end{equation}
where $Z^{n}$ corresponds to the replicated partition function. 
In the replica space, the average over the disorder may be evaluated and we 
adopt  the 1S-RSB; this procedure leads to the following free energy~\cite{MMN}:  
\begin{equation}
\begin{split}
 \beta f &
= \frac{(\beta J)^2}{4}x(q_1^2 -q_0^2)  - \frac{(\beta J)^{2}}{4} (1 - q_1)^{2} + \frac{\beta J_0}{2} m^2 
\\& - \frac{1}{x} \int Dz \ln\int Dv [2 \cosh \Xi (z,v)]^x, \\
\end{split}
\label{eq9}
\end{equation}
with $Dz \equiv  \frac{1}{\sqrt{2\pi}} dz  \exp (-z^{2}/2)~$ and 
all integrals should be considered over the whole interval (from  
$-\infty$ to $+\infty$). The 1S-RSB 
internal field $\Xi(z,v)$ is given by
\begin{equation}
\begin{split}
\Xi(z,v) &= \beta  J (\sqrt{q_0 +   \Theta (1 - n)  } z + \sqrt{q_1 - q_0}  v)
\\ & + \beta J_0m  + \beta ( H_l + n \Theta)~. 
\end{split}
\label{intfield}
\end{equation}
The equations above apply to both types of random fields, through the 
identifications $n=0$ and $\Theta = (\Delta/J)^2$ (Gaussian RF), whereas 
$n = 1$ and $\Theta = h_0$ (bimodal RF).   
It should be mentioned that, in the present work the results for the bimodal RF become
independent of $p$, which may be seen by means of a change of variables
$z \rightarrow -z \ $~\cite{Morais}. 
Therefore, the analysis of the bimodal distribution becomes completely 
equivalent to the one of a shifted uniform field, given by $H_l+h_{0}$.
Since the analysis of the SK model in the presence of a uniform field
has been carried in the literature by many 
authors~\cite{MMN,Morais,Pirc,Kim2002}, from now on we focus 
our analysis to the Gaussian random field, for which the internal 
field of Eq.~(\ref{intfield}) becomes  

\begin{equation}
\begin{split}
\Xi(z,v) &= \beta  J (\sqrt{q_0 +   (\Delta/J)^{2}} \, z + \sqrt{q_1 - q_0}  \, v)
\\ & + \beta J_0m  + \beta H_l~. 
\end{split}
\label{intfieldgauss}
\end{equation}

\noindent
The 1S-RSB parameters $q_0$, $q_1$ and $x$ should extremize 
the free energy of Eq.~(\ref{eq9}), from which the RS solution is 
recovered when $q=q_0=q_1~$ \cite{Pirc,Kim2002}. 
The linear susceptibility $\chi_1 = \frac{ \partial m}{ \partial H_l} |_{H_l \rightarrow 0 }$ 
is given by $\chi_1 = \beta[1 -q_1+ x(q_1-q_0)]$~\cite{Parisi80b} when $J_0=0$. 
The nonlinear susceptibility $\chi_3$ can be obtained 
from $\chi_3= -\frac{1}{3!} \frac{ \partial^3 m}{ \partial H_l^3} |_{H_l \rightarrow 0 }$.  
Moreover, important effects on $\chi_3$ appear already inside the region 
where the RS solution is stable, more precisely,  
$\chi_3$ presents a rounded maximum at a temperature 
$T^{*}$, above the SG transition.  
Particularly, we can expand $q$ and $m$ in powers of $H_l$, 
for $J_0=0$, as (following Wada~\cite{wada})

\begin{equation}
q(H_l)= Q_0 + Q_2 H_l^2, 
\end{equation}
\begin{equation}
m(H_l)= \chi_1 H_l + \chi_3 H_l^3,
\end{equation}

\noindent
which results in 

\begin{equation}
 \chi_3 (T) =  \frac{\beta ^3}{3} \left( 3  J^2 Q_2  + 1 \right) I_{0},
 \label{eqx_2}
\end{equation}

\noindent
with $Q_0 = \int Dz \tanh^2 \Xi_0(z)$ 
and $Q_2=  \frac{ \partial^2 q}{ \partial H_l^2} |_{H_l \rightarrow 0 }$,  
where the RS internal field is obtained from Eq.~(\ref{intfieldgauss}) by 
setting $H_l=0$ and 
$q_{0}=q_{1}=q$, i.e., $\Xi_0(z)= \beta  J (\sqrt{q +  (\Delta/J)^{2}} \, z)$ and
\begin{equation}
I_0 = \int Dz  
[ \mbox{sech}^4 \Xi_0(z) - 2 \ \mbox{sech}^2 \Xi_0 (z) 
\mbox{tanh}^2 \Xi_0(z) ]. \\
\label{eqfirst}
\end{equation}
Moreover, $Q_2$ can be obtained as
\begin{equation}
Q_2= \frac{1}{2!} \frac{\partial^2 q}{ \partial H_l^2}|_{H_l \rightarrow 0}  = \frac{\beta^2 I_0}{1 - (\beta J)^2 I_0}~,  
\label{eqder}
\end{equation}
so that $\chi_3(T)$ becomes
\begin{equation}
 \chi_3 (T) = - \frac{\beta^3}{3} \left[ \frac{ 3 (\beta J)^2I_0}{1 - (\beta J)^2 I_0} + 1 \right]I_{0}~. 
 \label{chi_gauss}
\end{equation}
These results hold when the RS solution is stable,
given by a positive value of the eigenvalue 
$\lambda_{\rm AT}$~\cite{SNA,NNCC,ANC,MMN,Morais},

\begin{equation}
\lambda_{\rm AT} = 1 - (\beta J)^2 \int Dz ~ \mbox{sech}^4 \Xi_0(z).
\label{lat}
\end{equation}

Particularly, $\chi_3(T)$ can be written in terms of 
$\lambda_{\rm AT}$,

\begin{equation}
\chi_3 (T) = \frac{\beta^3}{3} \left[ \frac{3}{\lambda_{\rm AT} + (\beta J)^2 I_1}   
- 2 \right] I_{0}~, 
\label{chigamma}
\end{equation}
where
\begin{equation}
I_1 = 2 \int Dz \, \mbox{sech}^2 \Xi_0(z) \mbox{tanh}^2 \Xi_0(z) .
\label{I1}
\end{equation}
In the absence of RFs, $I_1=0$ in the PM phase, implying on 
a divergence of $\chi_3$ 
when $\lambda_{\rm AT} = 0$, as expected~\cite{Binder}. 
Moreover, in the presence of RFs, one has that $I_1>0$,
so that Eq.~(\ref{chigamma}) leads to a rounded maximum at 
a temperature $T^{*}$. 

\section{Results and Discussion}

\begin{figure}[htb]
\includegraphics[angle=-90,width=\columnwidth]{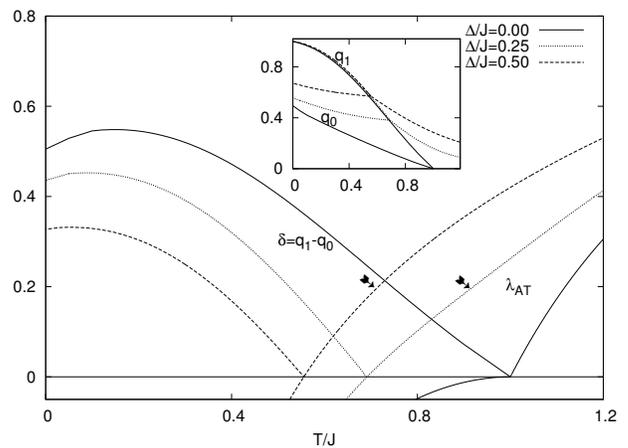}
\caption{The 1S-RSB parameter $\delta\equiv q_1-q_0$ and 
the eigenvalue $\lambda_{\rm AT}$ are presented 
versus the dimensionless temperature $T/J$, for typical values of 
$\Delta/J$. 
The inset shows the SG parameters
$q_1$ and $q_0$ separately versus the dimensionless temperature. 
The freezing temperature $T_f$ is identified with the onset of 
RSB, where $\lambda_{\rm AT}=0$, or equivalently, where the parameter
$\delta$ becomes nonzero. 
The arrows indicate the temperature $T^{*}$, where  
$\chi_3$ presents a rounded maximum, showing that $T^{*}>T_f$.
Due to the usual numerical difficulties, the 
low-temperature results [typically $(T/J)<0.05$)] correspond to 
smooth extrapolations from higher-temperature data.
}
\label{figop}
\end{figure}

Numerical results are now presented. The effects of RF's 
on the SG order parameters $q_0$, $q_1$, $\delta\equiv q_1-q_0$, 
susceptibilities $\chi_1$ and $\chi_3$, as well as  
the stability of the RS solution (i.e., $\lambda_{\rm AT}$) are discussed. 
In particular, 
the onset of RSB (location of $T_f$)  and how $\chi_3$
behaves in the neighborhood of the SG phase transition are studied. 

For instance, Fig.~\ref{figop} shows that the SG order-parameters behavior, 
signaling RSB ($\delta>0$), occurs at lower temperatures due to 
the presence of RFs, i.e.,  the increase of $\Delta/J$
moves $T_f$ to lower temperatures. 
The freezing temperature $T_f$, which is located within the 1S-RSB scheme 
as the onset of the parameter $\delta$, is shown herein to coincide 
with $\lambda_{\rm AT}=0$. As presented in the inset of 
Fig.~\ref{figop}, the RFs
induce the order parameters $q_0$ and $q_1$ for 
$T>T_f$, where the RS solution is stable [$q=q_0=q_1$, $\delta=0$ 
and $\lambda_{AT}>0$] characterizing the PM phase. 
In the cases $(\Delta/J)>0$ one notices that $T^{*}>T_f$, with the 
arrows indicating the temperature $T^{*}$ where  
$\chi_3$ presents a rounded maximum.   

\begin{figure}[htb]
\includegraphics[angle=-90,width=\columnwidth]{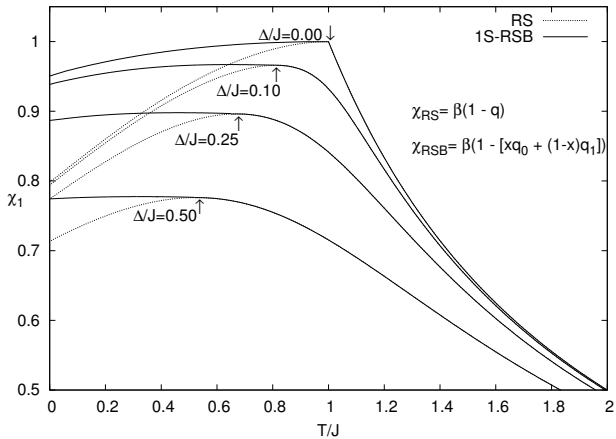} 
\caption{Magnetic susceptibility $\chi_1$ versus $T/J$ for different  
values of $\Delta/J$. The arrows indicate the onset of 
the RSB solution ($\lambda_{\rm AT}=0$), defining the temperature $T_{f}$. 
Below $T_{f}$, solid and dotted lines indicate linear susceptibilities computed using 
1S-RSB ($\chi_{\rm RSB}$) and 
RS ($\chi_{\rm RS}$) solutions, 
respectively. 
Due to the usual numerical difficulties, the 
low-temperature results [typically $(T/J)<0.05$)] correspond to 
smooth extrapolations from higher-temperature data.
}
\label{chi1}
\end{figure} 

As shown in Fig.~\ref{chi1}, the magnetic susceptibility 
$\chi_1$ exhibits a clear cusp at 
$T_f$ in the absence of the RF, whereas in the presence of a RF,
one notices a smooth behavior around $T_{f}$. Below this temperature, the
1S-RSB and RS solutions become distinct, with the former
presenting higher values, being weakly dependent on the
temperature. 

\begin{figure}[htb]
\includegraphics[angle=-90,width=\columnwidth]{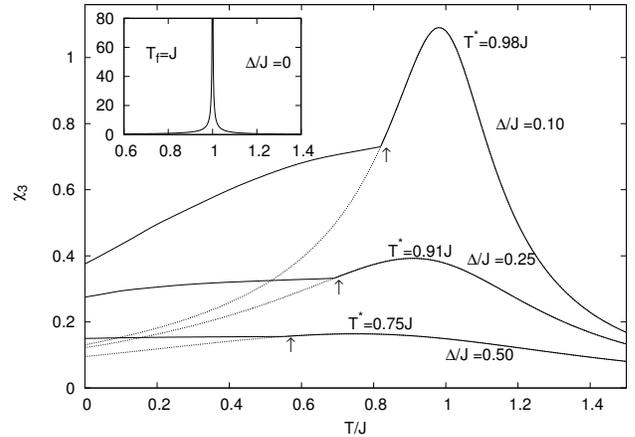} 
\caption{The susceptibility $\chi_3$ as a function of $T/J$ for different  
values of $\Delta/J$. The arrows indicate the onset of 
the RSB solution ($\lambda_{\rm AT}=0$), defining the temperature $T_{f}$. 
Below $T_{f}$, solid and dotted lines indicate 1S-RSB and RS solutions, 
respectively. The temperature $T^{*}$, where  
$\chi_3$ presents a rounded maximum, is estimated in each case 
shown. In the inset we exhibit the $\chi_3$ behavior without the RF. 
}
\label{chi3}
\end{figure}

In Fig.~\ref{chi3} we present results for the nonlinear susceptibility $\chi_3$, 
computed directly from the numerical derivatives  
$\chi_3= -\frac{1}{3!} \frac{ \partial^3 m}{ \partial H_l^3} |_{H_l \rightarrow 0 }$.  
As a check, for $T\geq T_f$, we verified that these results coincide with
those obtained from Eq.~(\ref{chigamma}). 
For the case without RFs, $\chi_3$ shows a strong divergence at $T_f$ 
(see inset in Fig.~\ref{chi3}); however, the presence of a  
RF eliminates this divergence, and rounded maxima appear in the 
$\chi_3$ curves, defining the temperature $T^{*}$ for each value of $\Delta/J$.
It is important to remark that $T^{*}$ is always higher than $T_f$.
Furthermore, the $T^{*}$ and $\chi_3$ values decrease for increasing 
values of $\Delta/J$.
Within the RSB region, similarly to what was shown for the linear 
susceptibility $\chi_1$
[cf. Fig.~\ref{chi1}],  $\chi_3$ also presents 
a split between the results with RS and 1S-RSB solutions.
However, differently from $\chi_1$, the nonlinear  susceptibility 
$\chi_3$ displays an evident discontinuity at $T_f$, when the 
1S-RSB solution is adopted. 

\begin{figure}[htb]
\includegraphics[angle=-90,width=\columnwidth]{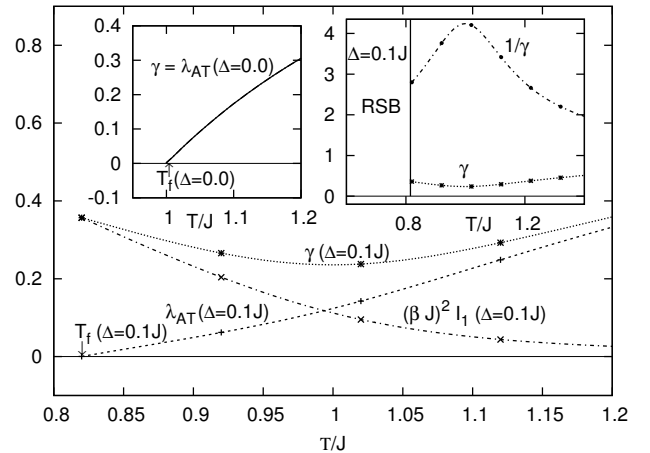} 
\caption{The quantities appearing in the denominator of Eq.~(\ref{chigamma}), 
$\gamma=\lambda_{\rm AT} + (\beta J)^2 I_1$,  are presented versus $T/J$, 
for $(\Delta/J)=0.1$. The arrows locate the freezing temperature $T_f$.  
The inset on the right shows in detail the behaviors of 
$\gamma$ and $1/\gamma$, 
for $(\Delta/J)=0.1$, in the region where the RS solution is
stable (to the left of this region one should use RSB); 
the quantity $1/\gamma$ presents a rounded maximum,
which is directly related with the one found in $\chi_3$. 
The inset on the left shows in detail the 
behaviors of $\gamma$ and $\lambda_{\rm AT}$, for $\Delta/J=0.0$,
which are responsible for the divergence of $\chi_3$ in the 
absence of RFs. 
}
\label{behI1G}
\end{figure}

An important quantity in Eq.~(\ref{chigamma}) is the denominator,
$\gamma=\lambda_{\rm AT} + (\beta J)^2 I_1$, which is illustrated 
in detail in Fig.~\ref{behI1G} versus $T/J$, for the typical
value $(\Delta/J)=0.1$.  
As a comparison, the inset on the left 
shows the behavior of 
$\gamma$ and $\lambda_{\rm AT}$, for $(\Delta/J)=0.0$; 
in this case, $\gamma=0$ leads to the divergence of 
$\chi_3$ in the absence of RFs. When $(\Delta/J) > 0$, 
one has that the contribution $(\beta J)^2 I_1>0$, so that
now $\gamma>0$.  
The two contributions,  
$\lambda_{\rm AT}$ (that increases for increasing values of $T/J$) and 
$(\beta J)^2 I_1$ (that decreases for increasing values of $T/J$), 
are presented separately, leading to a minimum
value for $\gamma$, which is found to occur very close to the 
temperature  $T^{*}$.
The inset on the right shows the maximum attained by $1/\gamma$, 
appearing inside the region where RS is stable; to the left
of this region, one should analyze these quantities within RSB. 
This maximum is directly related with the one presented in 
Fig.~\ref{chi3}, at the temperature $T^{*}$, and since this 
temperature is found in the RS region, we consider 
the rounded maximum to occur in the paramagnetic phase. 
One should remind the role played by the RF on the replicon,
leading to a shift in the freezing temperature towards lower
temperatures, i.e., 
$T_{f}(\Delta>0)<T_{f}(\Delta=0)$~\cite{SNA,MMN,Morais}.  
Hence, in Fig.~\ref{behI1G} one notices that in the temperature
range $T_{f}(\Delta>0)<T<T_{f}(\Delta=0)$ the behavior 
of the denominator $\gamma$ changes completely from decreasing  
to increasing. This inversion yields the minimum of $\gamma$, which  
is the ultimate mechanism leading to the rounded maximum of $\chi_3$ 
at $T^*$. 

\begin{figure}[htb]
\includegraphics[angle=-90,width=\columnwidth]{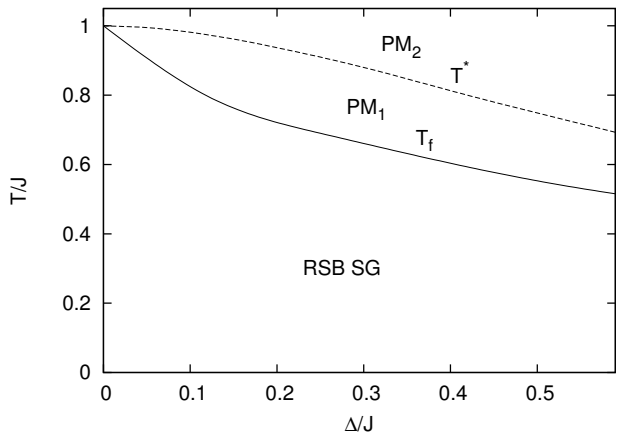} 
\caption{Phase diagram $T/J$ versus $\Delta/J$ showing the paramagnetic
and SG phases. 
The freezing temperature $T_{f}$, signaling the onset 
of RSB, defines the SG phase for $T<T_{f}$. 
For completeness, we also
present the line associated with the maximum of $\chi_{3}$, defining
the temperature $T^{*}$ (dashed line). The possibility of two paramagnetic 
phases (${\rm PM}_{1}$ and ${\rm PM}_{2}$) is discussed in the text. 
} 
\label{Tf_vs_T*}
\end{figure}

In Fig.~\ref{Tf_vs_T*} we present the phase diagram of the model,
showing the paramagnetic and SG phases. 
The SG phase is associated with the onset of RSB, being signaled 
by the zero of the replicon of Eq.~(\ref{lat}) (i.e., $\lambda_{\rm AT}=0$), 
which defines the freezing temperature $T_{f}$. The temperature $T_{f}$
is lowered due to the RFs; in fact, such a decrease in  
$T_f$ can be verified analytically for 
$(\Delta/J) \ll 1$, in  which case an expansion can be obtained from 
Eq.~(\ref{lat})~\cite{SNA,Morais}, 
\begin{equation}
{T_f \over J} \approx 1  -  \left({3 \over 4} \right)^{1/3} 
\left( {\Delta \over J} \right)^{2/3}~. 
\label{eXPtF}
\end{equation}
The dashed line in Fig.~\ref{Tf_vs_T*} represents the temperature $T^{*}$
($T^{*}>T_{f}$), characterizing the maximum of $\chi_{3}$,
which exists for any $(\Delta/J)>0$.
For $T>T^{*}$ the phase ${\rm PM}_{2}$ occurs, along which one has 
weak correlations and consequently, the usual paramagnetic type of 
behavior. However, close to  $T^{*}$, and particularly, for temperatures
in the range $T_{f}<T<T^{*}$, one expects a rather nontrivial behavior
in real systems, as happens with the compound 
LiHo$_x$Y$_{1-x}$F$_4$, resulting in very controversial 
interpretations~\cite{Schechter1,Laflorencie06,Gingras06,Gingras11,jonsson07,%
anconatorresprl2008,mydosh2015}.
Due to such aspects, herein we call the temperature region
$T_{f}<T<T^{*}$ of ${\rm PM}_{1}$. The line 
${\rm PM}_{1}$--${\rm PM}_{2}$ may not characterize a real phase 
transition, in the sense of a diverging $\chi_{3}$, but the region 
${\rm PM}_{1}$ is certainly characterized by a rather nontrivial dynamics.
As shown in Fig.~\ref{behI1G}, the region ${\rm PM}_{2}$ presents
small values for the quantity
$I_{1}$ [cf. the denominator of Eq.~(\ref{chigamma})], whereas along 
${\rm PM}_{1}$ the couplings between RFs and spins become 
dominant, as compared with thermal fluctuations and $I_1$ increases
significantly.  
As a possible relation, one should have a growth of free-energy barriers 
in the region ${\rm PM}_{1}$, leading to a slow dynamics, 
whereas only below $T_{f}$ the nontrivial ergodicity breaking appears, 
typical of RSB in SG systems.   
It is important to remind also that Griffiths singularities are found currently
in disordered magnetic systems, like for site-diluted 
ferromagnets~\cite{griffithssing}, as well for a ferromagnet in a random
field~\cite{dotsenkosing}. 
Whether the region ${\rm PM}_{1}$ in the 
present problem may be related to this later type of behavior concerns
a matter for further investigation. 

As already mentioned and addressed in several 
works~\cite{wada,suzuki,Chalupa77}, 
in the absence of RFs, the SG phase transition is given by the divergence 
of $\chi_3$ at $T_f$. In Parisi's mean field theory this divergence is 
directly related with the onset of RSB, signaled by a zero of the  
replicon $\lambda_{\rm AT}=0$~\cite{Binder}. 
However, the presence of RFs induce the SG order parameter $q$ 
in the PM phase, within the RS solution. Moreover, 
$\chi_3$ no longer diverges at the SG transition temperature, but instead,  
presents a rounded maximum at $T^*$, which becomes smoother 
as $\Delta/J$ increases. Such difference with respect to the case 
without RFs can be understood from Eqs.~(\ref{chigamma}) and~(\ref{I1}).
In fact, the term $I_1$ in Eq.~(\ref{chigamma}), which is responsible for these
effects, can be rewritten as
\begin{equation}
I_1 = 2(q - r),   
\label{I11}
\end{equation}
with
\begin{equation}
\langle S^{\alpha}S^{\beta} \rangle \equiv   q = \int Dz~ \mbox{tanh}^2 \Xi_0(z)~,  
\end{equation}
and
\begin{equation}
\langle S^{\alpha} S^{\beta}S^{\gamma}S^{\delta}\rangle  \equiv r = \int Dz~ \mbox{tanh}^4 \Xi_0(z)~.    
\end{equation}
These equations lead to 
$\gamma=\lambda_{\rm AT} + (\beta J)^2 I_1= 1- (\beta J)^2(1-4q +3 r)$,
which is precisely the longitudinal eigenvalue of the RS
stability analysis~\cite{Almeida78,Binder}.  
This longitudinal eigenvalue is related with the magnitude of the fluctuations  
of the of RS SG order parameter $q$.  
Hence, the maximum of $\chi_3$ at 
$T^*$ becomes completely unrelated with the SG phase transition 
when $\Delta$ departs from zero, being directly associated with the
longitudinal eigenvalue.    

\section{Conclusions}

The role of random fields on the 
spin-glass freezing temperature, as 
well as on the nonlinear susceptibility, was analyzed. For that, we have 
investigated the Sherrington-Kirkpatrick model in the presence of random
fields, following a Gaussian distribution characterized by a width $\Delta$,
within a one-step replica-symmetry-breaking procedure.  
We have shown that the divergence in $\chi_3$ only occurs in the
absence of random fields, and that $\chi_3$ exhibits a broad maximum 
at a temperature $T^{*}$ for $\Delta>0$.
The freezing temperature $T_f$ is associated with the onset 
of replica-symmetry-breaking, signaled by the zero of the 
Almeida-Thouless (replicon) eigenvalue, occurring at lower 
temperatures, i.e., $T^{*}>T_f$ for a given value of $\Delta$.  

The splitting between $T_f$ and $T^*$, for $\Delta>0$, was studied 
by analyzing the contribution due to the random fields in the replica 
symmetry spin-glass order parameter.
Particularly, we have shown that the behavior of $\chi_3$ is not regulated 
only by the spin-spin correlations associated to the Almeida-Thouless line,
but also to correlations coming from the longitudinal eigenvalue. 
These correlations play an important role inside the paramagnetic phase, 
when the random fields are applied, being  
responsible for the maximum in $\chi_3$, although they are not 
directly associated with the spin-glass phase transition. 

Although the present results refer a specific model, we expect they 
could shed some light in the theoretical and experimental 
description of disordered magnetic systems like, for instance, 
the compound LiHo$_x$Y$_{1-x}$F$_4$. 
Considering recent observations in  
LiHo$_x$Y$_{1-x}$F$_4$, we follow the proposal that 
an applied transverse field $B_t$
induces longitudinal random fields~\cite{Gingras06}, and thus, 
we assume herein $\Delta=\Delta(B_t)$. 
In this way, one can interpret 
the present results, e.g., the temperatures $T^{*}$ and $T_f$, as 
manifestations of the transverse field. Based on this, we point out below
two possibilities, which may contribute to elucidate the 
recent controversies on this 
system~\cite{Schechter1,Laflorencie06,Gingras06,Gingras11,jonsson07,%
anconatorresprl2008,mydosh2015}.  (i) The temperature $T^{*}$ associated with the rounded maximum in the
nonlinear susceptibility does not signal any phase transition, being an effect of random fields inside the paramagnetic phase,  
although
it is related to a minimum of the longitudinal eigenvalue, and
hence, to large fluctuations in the replica-symmetric spin-glass
order parameter. A true spin-glass phase transition,  indicated through the Sherrington-Kirkpatrick model, by means of the Almeida-Thouless line, should   
  occur at the lower temperature $T_f$. (ii) There is no spin-glass phase transition in LiHo$_x$Y$_{1-x}$F$_4$  
in the presence of a transverse field, 
implying that the replica-symmetry-breaking procedure does not apply 
to this compound.   
Certainly, these two points require meticulous experimental observations 
for temperatures around $T^{*}$, which has been the most 
investigated temperature region so far, as well as 
below $T^{*}$, representing a challenge for experiments. 

\section*{Acknowledgments}

This work was supported by the Brazilian funding agencies: CNPq processes 306720/2013-2,  474559/2013-0 ,  443565/2014-7 and CAPES.

\end{document}